\documentclass{aa}  
\usepackage{graphicx}
\usepackage{txfonts}
\usepackage[figuresright]{rotating}
\begin{document}
   \title{Atmospheric coherence times in interferometry: definition and measurement}

   \subtitle{}

   \author{A. Kellerer
          \inst{1}
          \and
          A. Tokovinin\inst{2}
          }
   \offprints{A. Kellerer}

   \institute{European Southern Observatory, 
Karl-Schwarzschild-Strasse 2, 85748 
Garching bei M\"unchen, Germany \\
              \email{akellere@eso.org}
         \and
             Cerro Tololo Inter-American Observatory, Casilla~603, La Serena, 
Chile \\
             \email{atokovinin@ctio.noao.edu}
             }

   \date{Received June 09, 2006; accepted  }

 
  \abstract
   {}
   {Current  and  future  ground-based  interferometers
require knowledge  of the atmospheric time constant  $t_0$, but this
parameter has diverse definitions. 
Moreover, adequate techniques for monitoring $t_0$ still have to be implemented. }
   {We derive  a new formula for the structure function
of the fringe phase (piston) in a long-baseline interferometer,
and review  available  techniques  for  measuring  the atmospheric time constant 
and  the shortcomings.}
   { It is shown  that  the standard  adaptive-optics  atmospheric time  constant
is sufficient for quantifying  the  piston coherence  time, with only
minor modifications. The residual  error of a fast fringe tracker
and  the loss  of  fringe visibility  in  a finite  exposure time  are
calculated in terms of the same parameter. 
A new method based  on the fast variations of defocus is
proposed. The formula for relating the  defocus speed to the time constant
is  derived.   Simulations  of  a  35-cm   telescope  demonstrate  the
feasibility of this new technique for site testing.}
   {}

   \keywords{ atmospheric effects,
   instrumentation: interferometers,
   site testing
               }

   \maketitle
%

\section{Introduction}

Astronomical  sites for  classical observations  are  characterized in
terms  of  atmospheric  image   quality  (seeing).  For  high-angular
resolution techniques such as adaptive optics (AO) and interferometry,
we need to know  additional parameters. The atmospheric coherence time
is one of these. Here  we refine the definition of the interferometric
coherence time,  review available techniques, and propose  a new method
for its measurements.

The AO  {\it time constant,  $\tau_0$,\/} is a well-defined parameter
related  to the  vertical distribution  of turbulence  and  wind speed
(Roddier \cite{Roddier81}).   To correct  wave fronts in real  time, a
sufficient number of photons from the guide star is needed within each coherence
area during time $\tau_0$.  This severely restricts the
choice of natural  guide stars and tends to impose  the complex use of
laser guide stars (Hardy \cite{Hardy}).  It is shown below that new,
simple methods of $\tau_0$ monitoring are still needed.

Modern ground-based stellar interferometers attain extreme resolution,
but their sensitivity is limited  by the atmosphere.  Even at the best
observing sites, such as Paranal in Chile, fast fringe tracking is not
fully  operative yet,  and one therefore tends to  employ  exposure
times that are short  enough to ``freeze" the atmospheric turbulence.
The price is a substantial loss in limiting magnitude. It is hence
important to  measure the {\it time constant, $t_0$\/}, of the piston 
-- i. e. the mean phase over the telescope aperture -- at  existing and
future sites.
However, the exact definition of $t_0$ is not clear, any
more than are methods to measure it.  Do we need an interferometer to
evaluate $t_0$?  Is $t_0$ different from $\tau_0$?  Does it depend on the
aperture  size and  baseline?  
We  review various  definitions of  the interferometric  time constant
based on the piston {\it structure function\/} (SF), on the error of a  fringe  tracker,
and on the loss of fringe contrast during a finite exposure time. 
It  is  shown that  the  piston  time constant  is
proportional to the AO coherence  time $\tau_0$, both depending on the
same combination  of atmospheric parameters.  

During  site exploration  campaigns, one
would   like   to  predict   the   performance   of  large   base-line
interferometers, and it is desirable  to do this with single-dish and,
preferably, small telescopes. The existing techniques
for $\tau_0$ measurement are listed and a new method for site
testing proposed.

\section{Atmospheric coherence time in interferometry}
\label{fringe_motion}

\subsection{Atmospheric coherence time $\tau_0$}
\label{sec:heuristic}

First,  we introduce the relevant atmospheric  parameters and  the AO
time constant  $\tau_0$.  For convenience, we outline the essential
formulae, but for the general background, we  refer the reader
to Roddier (\cite{Roddier81}).

The spatial and temporal fluctuations of atmospheric phase distortion
$\varphi$ are usually described by the SF
\begin{equation} 
D_{\varphi}({\bf r},t) = \langle \left[ \varphi({\bf
    r'},t')-\varphi({\bf r} +{\bf r'}, t + t') \right]^2 \rangle,
\label{eq:D-def}
\end{equation}
which depends on the transverse  spatial coordinate ${\bf r}$ and time
interval $t$.  The angular brackets indicate statistical average.

The atmosphere consists  of many layers.  The contribution  of a layer  $i$ of thickness d$h$
at altitude $h$ 
to the turbulence intensity is specified in terms of $C_n^2(h) {\rm
d}h$,   equivalently    expressed   through   the  {\it Fried   parameter\/}
$r_{0,i}^{-5/3} =  0.423 k^2 C_n^2(h)  {\rm d}h$, $  k = 2  \pi /\lambda$
being the wavenumber. 
The spatial SF in the {\it inertial range\/} (between
inner and outer scales) is
\begin{equation} 
D_{\varphi}({\bf r},0) = 6.883 \; (|{\bf r}|/r_0)^{5/3} .
\label{eq:D-Fried}
\end{equation} 

It  is assumed  that each  layer moves  as a  whole with  the velocity
vector  ${\bf V}(h)$  (Taylor  hypothesis).  
The  temporal  SF of  the
piston fluctuations $D_{\varphi,i}(0,t)$ in  one small aperture due to
a single layer is then equal to,  the spatial SF at shift $V t$,
\begin{equation} 
D_{\varphi,i}(0,t) = 6.883 \; [V(h) t/r_{0,i} ]^{5/3}.
\label{eq:Dphi}
\end{equation}
Summing the contributions of all layers, we obtain
\begin{eqnarray} 
D_{\varphi}(0,t) &=& 2.910 \; t^{5/3} \;  k^2  \; 
\int_0^{+\infty} V^{5/3}(h) C_n^2(h)  {\rm  d}h \nonumber \\
&=& 6.883 \; ( t  \overline{V}_{5/3} /r_0)^{5/3} 
= (t/\tau_0)^{5/3},
\label{eq:Dphi2}
\end{eqnarray}
where $\tau_0  = 0.314 \; (r_0  / \overline{V}_{5/3})$ is  the AO time
constant (Roddier 1981) and  the average wind speed $\overline{V}_p$
is computed as
\begin{equation}
\overline{V}_p  =  \left[ 
\frac{\int_0^{+\infty} V^{p}(h) C_n^2(h)  {\rm  d}h  }
{\int_0^{+\infty}  C_n^2(h)  {\rm  d}h  } 
\right] ^{1/p} .
\label{eq:Vp}
\end{equation}
The  formulae are  valid  for  observations at  zenith. At angle $\gamma$
from the zenith, the optical path  is increased in proportion to the
{\it air  mass\/}, $\sec  \gamma$, and the  SF increases  by the  same  factor. 
Further, the transverse component of the wind velocity changes.
In the  following, we
neglect these complications and  consider only observations at zenith,
but the analysis of real data must account for $\gamma \neq 0$.

\subsection{Piston time constant}
\label{sec:t0}

In an interferometer with a large baseline 
($B \gg L_0$, where $L_0$: turbulence outer scale)
the phase patterns over
the  apertures are  uncorrelated on  short time  scales.  Thus,  for a
small  time interval  ($t < B/V$), the  SF of  the  phase difference 
$\phi$ (do not confuse with the  phase $\varphi$)  in an
interferometer  with two  small  apertures will simply be two  times
larger, $D_\phi(t) = 2 D_\varphi (0,t)$ (Conan et al. \cite{Conan95}).   
As a result the
differential piston variance reaches  1 rad$^2$ for a time  delay $t_0 =2^{-3/5}
\; \tau_0 = 0.66 \; \tau_0$.
Note that, in the case of smaller baselines and large outer scales
-- when the assumption $B \gg L_0$ becomes invalid -- $D_\phi(t) < 2 D_\varphi (0,t)$ 
and the resulting coherence time, accordingly, lies between $0.66 \; \tau_0$ and $\tau_0$. 
Yet, $B \gg L_0$ applies to the characterization of large baseline interferometers at low-turbulence sites.

When  an interferometer  with larger circular  apertures of  diameter  $d$ is
considered, phase fluctuations are  averaged inside each aperture.  As
shown later, for time increments smaller than $d/V$, the piston
structure function  is quadratic in $t$ and  is essentially determined
by the average wave-front tilt  over the aperture. The variance of the
gradient   tilt   $\alpha$   (in   radians)  in   one   direction   is
(Roddier \cite{Roddier81}, Conan et al. \cite{Conan95}, Sasiela \cite{Sasiela})
\begin{equation}
\label{tilt}
\sigma_{\alpha}^2 = 0.170 \;  \lambda^2 r_0^{-5/3} d^{-1/3} .
\end{equation}
We write the piston SF in this  regime as $D_\phi (t) \approx 2 \; (k
\sigma_{\alpha} V  t)^2$, sum the contributions of  all layers, and
obtain the expression
\begin{equation} 
D_{\phi}(t) \approx 
13.42 \; (\overline{V}_2 t / r_0 )^2 (r_0/d)^{1/3} = (t/t_1)^2 ,
\label{eq:Dphi3}
\end{equation}
where the modified time constant $t_1 = 0.273 \; (r_0/ \overline{V}_2)
\;  (d/r_0)^{1/6}$.  The  analysis of  the tilt  variance  with finite
outer scale  by Conan et  al.  (\cite{Conan2000}) is  applicable here.
The finite  outer scale  reduces the amplitude  of the tilt  and hence
increases the  piston time  constant, but this  effect depends  on the
aperture size and is not very strong for $d< 1$\,m.

Note that for  small time intervals there is a weak  dependence of the SF
on  the  aperture diameter.   Also,  the  wind  velocity averaging  is
slightly  modified.   However, the  expressions  for  $t_1$ and  $t_0$
produce  similar numerical  results  as  long as  $d/r_0$  is not  too
large. Thus, the system-independent definition of the AO time constant
(\ref{eq:Dphi2}) also gives  a  good  description  of  the  temporal
variations of the piston.

For time delays of  approximately  $B/V$ and larger,  the pistons  on two
apertures are no longer independent.   However, 
estimates of the time interval over which the Taylor hypothesis is valid 
range from $\sim 40$\,ms (Schoeck \& Spillar \cite{Schoeck}) 
to several seconds (Colavita  et al. \cite{Colavita87}).
Hence, at time intervals of 1\,s or more,  the Taylor hypothesis is insecure.
Moreover,  the finite
turbulence outer scale reduces the amplitude of slow piston variations
substantially.  Here  we concentrate  only on rapid  piston variations
where our approximations are valid.

\subsection{Piston power spectrum and structure function }
\begin{figure}
\centering
 \includegraphics[width=0.5\textwidth]{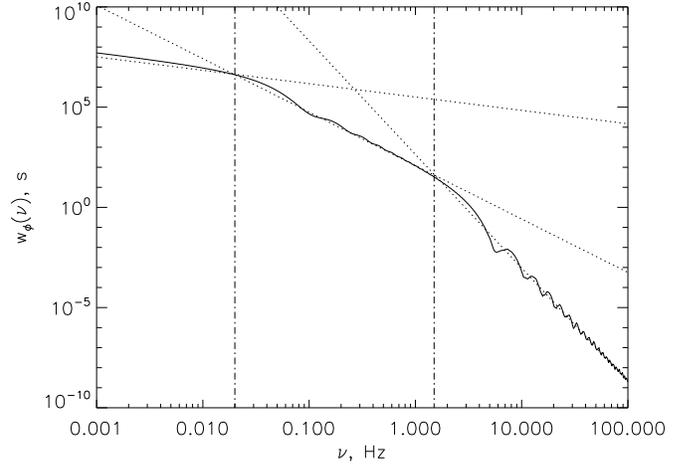}
\caption{Theoretical temporal  power spectrum of the  fringe position at
  0.5\,$\mu$m wavelength.  The two  telescopes are separated by 100\,m
  and  have  mirrors of  2\,m  diameter,  the  Fried parameter  equals
  $r_0$=11\,cm, the wind vector makes  an angle of $\alpha = 45^\circ$
  with the baseline,  $V = 10$\,m/s. The vertical  lines correspond to
  the frequencies: $0.2\,V /B$ and $0.3\,V/d$. The asymptotic power laws
  are  $\nu^{-2/3}$,  $\nu^{-8/3}$,   $\nu^{-17/3}$ from  lowest  to
  highest frequencies.}
\label{piston}
\end{figure}
\begin{figure}
\centering
 \includegraphics[width=0.5\textwidth]{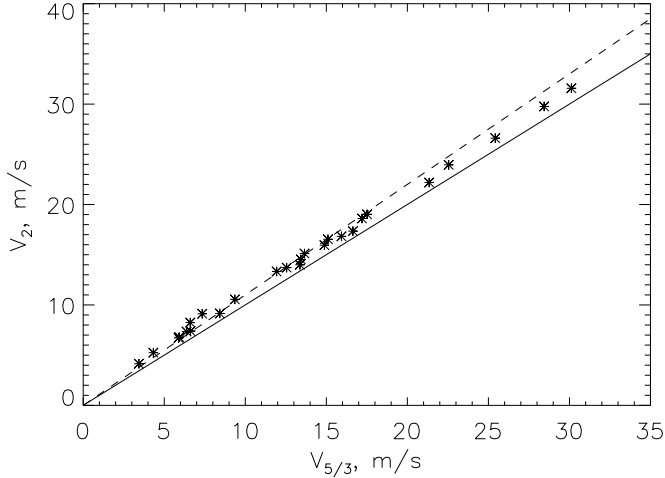}
\caption{Relation between average wind velocities $\overline{V}_{5/3}$
and  $\overline{V}_{2}$ for  26 balloon  profiles at  Cerro  Pachon in
Chile (Avila et al. \cite{Gemini}). The full line corresponds to equality, the dashed line is
$\overline{V}_{2} = 1.1 \; \overline{V}_{5/3} $.
\label{fig:v53} }
\end{figure}
\begin{figure}
\centering
 \includegraphics[width=0.5\textwidth]{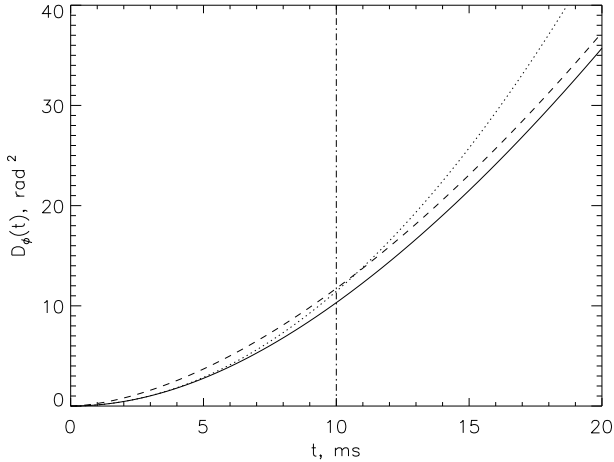}
\caption{Structure function of the fringe position for an interferometer with mirror diameters $d = 0.1$\,m,
$r_0 = 11$\,cm, $V = 10$\,m/s. The vertical  line corresponds to $t = d/V$. 
For $t < d/V$, the SF is quadratic in t (dotted line), cf. Eq.\,\ref{eq:Dphi3}. 
For longer time scales, $D_\phi \approx\,2 D_\varphi$ (dashed line).}
\label{fig:SF}
\end{figure}

The temporal  power spectrum of the atmospheric fringe position  has been
derived by Conan et  al. (\cite{Conan95}).  Their result is reproduced
in Appendix\,A with minor changes.  The temporal piston power spectrum
(\ref{eq:w-nu}) produced by a single turbulent layer is represented in
Fig.\,\ref{piston} for  a specific set of parameters.   
Because of the
infinite  outer  scale  $L_0$,  this  example  is  not  realistic  for
frequencies  below  $\sim 1$\,Hz.
Moreover, as  discussed  in Sect.\,\ref{sec:t0}, Taylor's frozen flow hypothesis 
becomes invalid at low frequencies.
Due to the infinite $L_0$, the asymptotic behavior of the spectrum, and 
in particular the cut-off frequencies, do not depend on the 
wind direction (Conan  et al. \cite{Conan95}), 
whereas, in the real case of a finite outer scale,
the cut-off frequencies are affected by wind direction, 
as described by Avila et al. (\cite{Avila97}).
Conan  et al.   (\cite{Conan95})  point out
that changing  turbulence intensity and wind speed  shift the spectrum
vertically and horizontally,  respectively, without changing the shape
of  the curve on  the log-log  plot.
In observations with a  small baseline ($\sim12$\,m), the proportionality
to  $\nu^{-2/3}$ at  low  frequencies and  to  $\nu^{-8/3}$ at  medium
frequencies  has   actually  been  measured,  e.g.    by  Colavita  et
al.  (\cite{Colavita87}).

Based on  the piston power spectrum,  we derive in Appendix\,A the
new  expression  of the  piston  SF valid  for  time  increments 
$t<  min(B/ \overline{V} , L_0/ \overline{V} )$:
\begin{equation}
D_{\phi} (t) \approx 13.76 \; (\overline{V} t /r_0)^2 \;
[ 1.17 \; (d/r_0)^2 + (\overline{V} t /r_0)^2 ]^{-1/6} .
\label{eq:Dphi-model}
\end{equation}
As seen in Fig.\,\ref{fig:SF}, for $t >  d/\overline{V}$, the piston averaging over  apertures is not
important  and we  obtain $D_\phi  =  2 D_\varphi$  in agreement  with
heuristic arguments.   For very short increments $  t \ll d/\overline{V}$,
(\ref{eq:Dphi-model}) reduces  to (\ref{eq:Dphi3}).  The  average wind
speed    is   $\overline{V}    \approx    \overline{V}_{5/3}   \approx
\overline{V}_{2}$.   The difference  between  $\overline{V}_{5/3}$ and
$\overline{V}_{2}$ is indeed small (Fig.\,\ref{fig:v53}).

\subsection{Error of a fringe tracking servo}
\label{sec:servo}
\begin{figure}
\centering
 \includegraphics[width=0.5\textwidth]{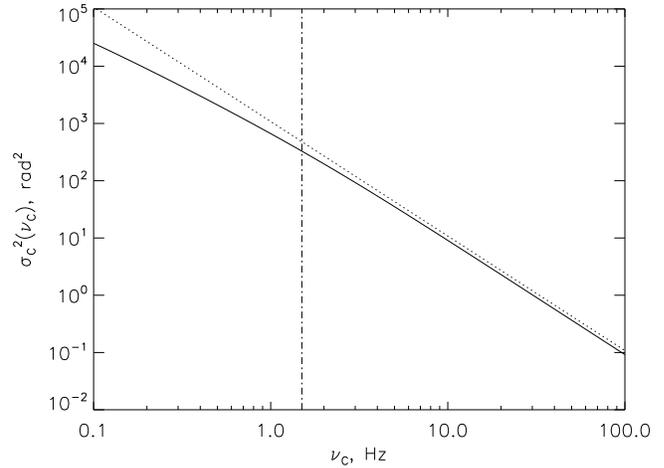}
\caption{Variance  of corrected  fringe  position as  a  function of  the
bandwidth frequency  of the correction  system. The parameters  of the
simulation are  identical to those  of Fig.\,\ref{piston}. 
At frequencies higher than $\nu_c = 0.3 V/d $ (vertical line), the variance
is approximated by $(2\pi\nu_ct_1)^{-2}$ (dotted line). }
\label{residual_variance}
\end{figure}

 A  fringe tracker  measures the  position of  the central  fringe and
computes a correction.  The  actual compensation equals the integrated
corrections applied  after each iteration. Our analysis  is similar to
the  classical work  by  Greenwood \&  Fried (\cite{Greenwood}).   
For a more detailed model that takes  the effect
of the finite exposure and response times of the phasing device into account, 
see the work by Conan et al.\,(\cite{Conan2000b}).
The error transfer function of a first-order phase-tracking loop equals
\begin{eqnarray}
T(\nu)&=&i\nu/(\nu_c+i\nu) ,
\end{eqnarray}
where  $\nu_c$ is  the 3\,dB  bandwidth of  the system.   The temporal
power  spectrum of  the corrected  fringe position  is $w_c(\nu)  = |T(\nu)|^2
w_\phi(\nu)$.  
The  residual piston  variance  characterizes the  performance of  the
phasing device. This variance is shown in Fig.\,\ref{residual_variance}
as a function of $\nu_c$ and is given by
\begin{eqnarray}
\label{eq:sigma}
\sigma_c^2(\nu_c)&=&\int_{-\infty}^{+\infty}\nu^2/(\nu_c^2+\nu^2) \; w_\phi (\nu) {\rm
  d}\nu .
\end{eqnarray}

When $v_c  < 0.3  V/d$, the fringe  tracker is  too slow and  leaves a
large residual error;  only fast trackers with $v_c >  0.3 V/d$ are of any 
practical interest.   In this case,  the dominant contribution  to the
residual  variance in  (\ref{eq:sigma}) comes  from  the frequencies
just  below $0.3 V/d$,  where the  filter is  approximated as  $(\nu /
\nu_c)^2$. Hence the residual variance is proportional to the variance
of  the piston  velocity.   There  is a  simple  relation between  the
residual error of the fringe tracker and the structure function of the
piston. For small arguments $t$, we can  replace $ 2 [ 1 - \cos( 2 \pi
\nu t)] \approx (2 \pi \nu t)^2$ in the expression (\ref{eq:Dphi4}) for
the phase  SF. Then the residual  error of the fast  fringe tracker is
simply
\begin{eqnarray}
\label{eq:sigma2}
\sigma_{c}^2(\nu_c)& \approx D_{\phi} [ 1 / (2 \pi \nu_c ) ]  
\approx (2 \pi \nu_c t_1)^{-2} .
\end{eqnarray}
Thus, we have established that the error of the fast fringe tracker and the
initial quadratic part of the piston SF are essentially determined by
the variance of piston velocity which, in turn, depends on the tilt
variance and the average wind speed $\overline{V}_2$.

\subsection{Summary of definitions and discussion}

Table\,\ref{tab:def} assembles different  definitions of the atmospheric
coherence time.  We have demonstrated  that the time constant $t_0$ of
the piston  SF is proportional to  the AO time  constant $\tau_0$. For
small time increments, a slightly modified parameter $t_1$ should be used.

A  different,  but  essentially  equivalent, definition  of  the  piston
coherence time $T_0 = 0.81 \; r_0 / \overline{V}_{5/3} = 2.58 \; \tau_0$
has been  given by  Tango \& Twiss (\cite{Tango80}) and  reproduced by
Colavita et al. (\cite{Colavita87}).  It is the integration time during
which the  piston variance  equals 1 rad$^2$. When fringes are integrated
over a time $T_0$, the mean decrease in squared visibility equals $1/e$.
Here we use  the more convenient definition $t_0  = 0.66 \; \tau_0$ based
on  the  temporal SF  and  warn  against  confusion with  Tango's
$T_0$. The  definition of  $T_0$ is  valid only for  $T >  d/V$, while
shorter integration times are of practical interest (see below).

The   performance   of   the   fringe-tracker   in   a   long-baseline
interferometer can  be characterized by the  atmospheric time constant
$t_1$ or,  equivalently, by  the average wind  speed $\overline{V}_2$.
The AO time constant $\tau_0$ (or $\overline{V}_{5/3}$) is also a good
estimator of the piston coherence time, especially for small apertures
$d \sim r_0$. 

\begin{table*}[ht]
\caption{Definitions of atmospheric time constants}
\label{tab:def}
\centering
\begin{tabular}{lll}
\hline
Quantity of interest & Formula & Time constant 
\medskip \\
\hline
\noalign{\smallskip}
Phase SF & $D_\varphi(t) = (t/\tau_0)^{5/3}$ & 
$\tau_0 = 0.314\; r_0/\overline{V}_{5/3}$  \smallskip \\
Piston SF, $t<  d/V$ & $D_\phi(t) = (t/t_1)^{2}$ & 
$t_1 = 0.273\; (r_0/\overline{V}_{2})(d/r_0)^{1/6}$  \smallskip \\
Piston SF,  $t> d/V$ & $D_\phi(t) = (t/t_0)^{5/3}$ & 
$t_0 = 0.66 \; \tau_0  $  \smallskip \\ 
Piston variance during an exposure $t> d/V$ & $\sigma^2_\phi (t)=
(t/T_0)^{5/3}$ & 
$ T_0 = 2.58\; \tau_0  $ \smallskip \\
Phase tracker error, $\nu_c> 0.3 \; d/V$ & $\sigma_c^2 (\nu_c)= (2 \pi \nu_c t_1)^{-2}$
& $t_1$  \smallskip \\
\hline
\end{tabular}
\end{table*}

In order to  reach a good magnitude limit,  all modern interferometers
have  large  apertures $d>r_0$.   The  atmospheric  variance over  the
aperture is $1.03\;(d/r_0)^{5/3} > 1$\,rad$^2$ and has to be corrected by
some means (tip-tilt guiding, full AO correction, spatial filtering of
the  PSF)  even  at  short  integration times.   The  temporal  piston
variance  will   also  be  $>$1\,rad$^2$   on time  scales of approximately
$r_0/\overline{V}$ and  longer. 
Hence exposure times shorter than $r_0/ \overline{V}$ or fast fringe
trackers are required in order to  maintain high fringe
contrast.  In this  regime, the relevant time constant 
that determines the visibility loss is $t_1$, rather than $\tau_0$ and $T_0$.

All definitions of atmospheric time constants contain a combination of
$r_0$ and  $\overline{V}$.  As  turbulence becomes stronger,  the time
constant  decreases, although  the  wind speed  may remain  unchanged.
Being  less correlated,  the parameters  $r_0, \overline{V}$  are thus
more  suitable   for  characterizing  atmospheric   turbulence  than  the
parameters  $r_0,  \tau_0$.    Astronomical  sites  with  ``slow''  or
``fast''  seeing should be  ranked in  terms of  $\overline{V}$ rather
than $\tau_0$.  A fair correlation between $\overline{V}$ and the wind
speed  at 200\,mB  altitude has  been  noted by  Sarazin \&  Tokovinin
(\cite{Sarazin2002}).

\section{Measuring the atmospheric  time constant}
\subsection{Existing methods of $\tau_0$ measurement}

\begin{table*}
\caption{Methods of $\tau_0$ measurement}
\label{tab:tau}
\centering
\begin{tabular}{ll l ll }
\hline
Method & Measurables & $d$, m & Problems  & Reference \smallskip \\
\hline \noalign{\smallskip}
SCIDAR & $C_n^2(h)$, $V(h)$ & $>$1 &  Needs large telescope  & Fuchs et al. \cite{SCIDAR} \\ 
Balloons & $C_n^2(h)$, $V(h)$ & none &  Expensive, no monitoring & Azouit \& Vernin \cite{Azouit} \\
AO system & $r_0$, $\tau_0$  & $>$1 &    Needs working AO & Fusco et al. \cite{Fusco2004} \\
SSS & $C_n^2(h)$, $V(h)$ & $>$0.4 & Low height resolution & Habib et al. \cite{Habib} \\
GSM & $r_0$, $V$, $\tau_{AA}$ & 4x0.1 & No obvious relation to $\tau_0$ and $t_1$ & Ziad et al \cite{GSM} \\
MASS      & $\tau_0^*$ & 0.02 & Biased (low layers ignored) & Kornilov et al. \cite{MASS1} \\
DIMM      & $r_0$  & 0.25 &  Indirect $\tau_0$ estimate & Sarazin \& Tokovinin \cite{Sarazin2002} \\
FADE  & $r_0$, $t_1$ & 0.35 & New method & This work \\
\hline
\end{tabular}
\end{table*}

Table\,\ref{tab:tau} lists methods  available for measuring the atmospheric
coherence  time $\tau_0$  or related  parameters. The 3rd
column gives an indicative diameter of the telescope aperture required for each method.
Short comments on each technique are given below.

SCIDAR (SCIntillation Detection And Ranging) has provided  good results 
on $\tau_0$. It  is not suitable for
monitoring because  manual data processing is still  needed to extract
$V(h)$,  despite efforts  to automate  the process.   Balloons provide
only    single-shot   profiles    of   low    individual   statistical
significance. The AO systems and interferometers give reliable results,
but are not suitable for testing new sites or for long-term monitoring.

The methods listed in the  next four rows of Table\,\ref{tab:tau} all
require  small  telescopes and  can  thus  be  used for  site-testing.
However, all  these techniques have some intrinsic  problems.  
SSS (Single Star SCIDAR) essentially extends the SCIDAR technique to small telescopes:
profiles of $C_n^2 (h)$ and $V(h)$ are obtained with lower height 
resolution than with the SCIDAR, and are then used to derive the coherence time.
The GSM (Generalized Seeing Monitor) can only  measure velocities of  
prominent layers after  careful data processing. 
A coherence time, $\tau_{AA}$ -- 
which, however, does not have a similar dependence on the turbulence 
profile than $\tau_0$ and $t_1$ --
is deduced from the angle of arrival fluctuations. 
MASS (Multi-Aperture Scintillation Sensor) is  a recent,  but  already well-proven,
turbulence monitor.   One of its observables  related to scintillation
in  a  2\,cm  aperture  approximates  $\overline{V}_{5/3}$  (Tokovinin
\cite{MASS}), but this averaging does  not include low layers and thus
gives a biased  estimate of $\tau_0$.  An even  less secure evaluation
of $\tau_0$ can be obtained  from DIMM (Differential Image Motion Monitor) 
by combining the measured $r_0$
with  meteorological data  on  the wind  speed  (Sarazin \&  Tokovinin
\cite{Sarazin2002}).

We conclude from this brief survey that a correct yet simple technique
for measuring   $\tau_0$  with  a  small-aperture   telescope  is  still
lacking. Such a method is proposed in the next section.

\subsection{The new method: FADE}
\begin{figure}
\centering
 \includegraphics[width=0.5\textwidth]{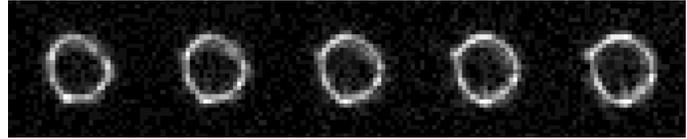}
\caption{Five  consecutive  ring images  distorted  by turbulence  and
  detector noise.  Each image  is 16x16 pixels ($13.8''$), the average
  ring  radius is  $3''$, the  interval between  images is  3\,ms, the wind
  speed is 10\,m/s.
\label{fig:movie} }
\end{figure}
\begin{figure*}
\centerline{
 \includegraphics[width=0.5\textwidth]{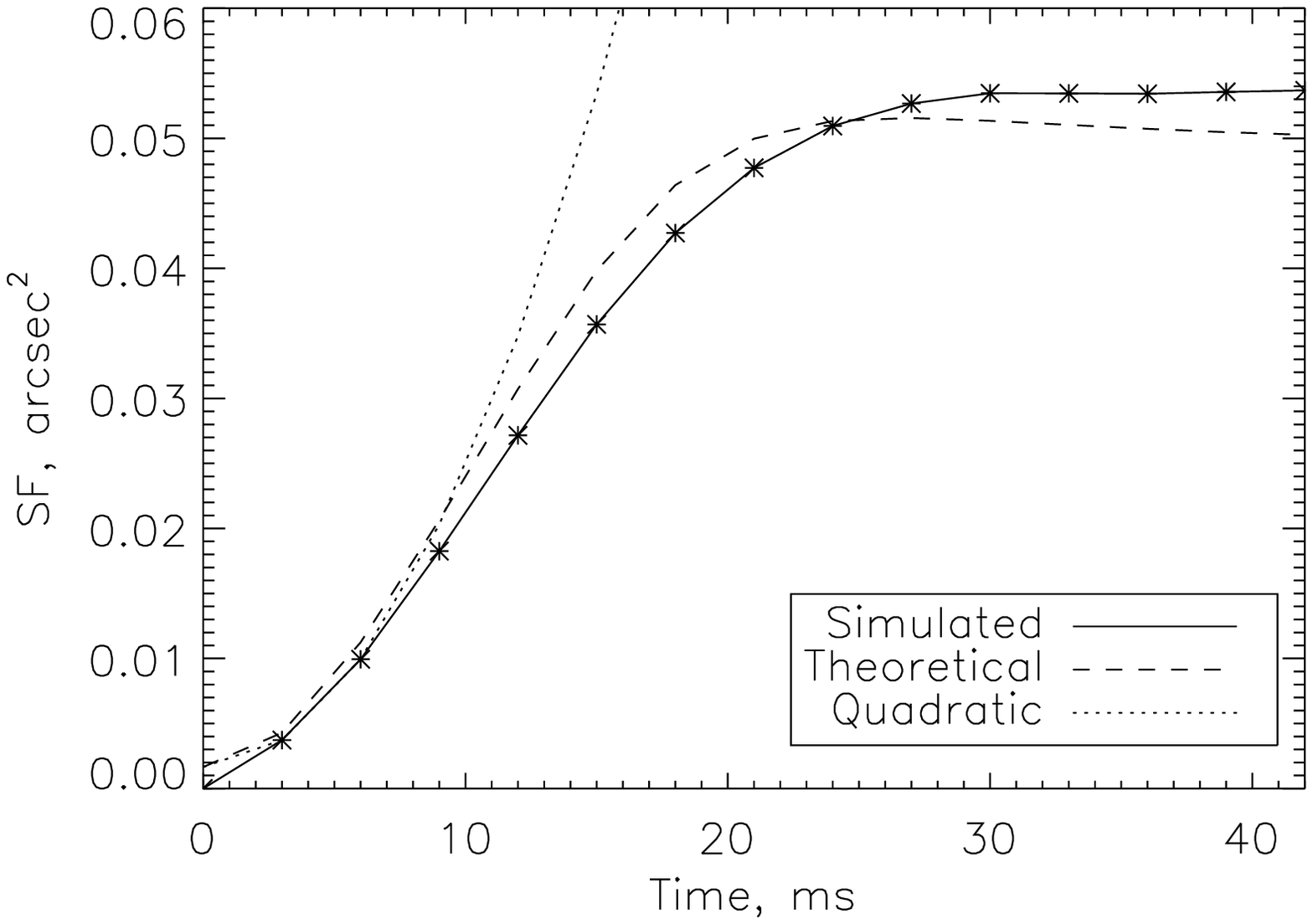}
 \includegraphics[width=0.5\textwidth]{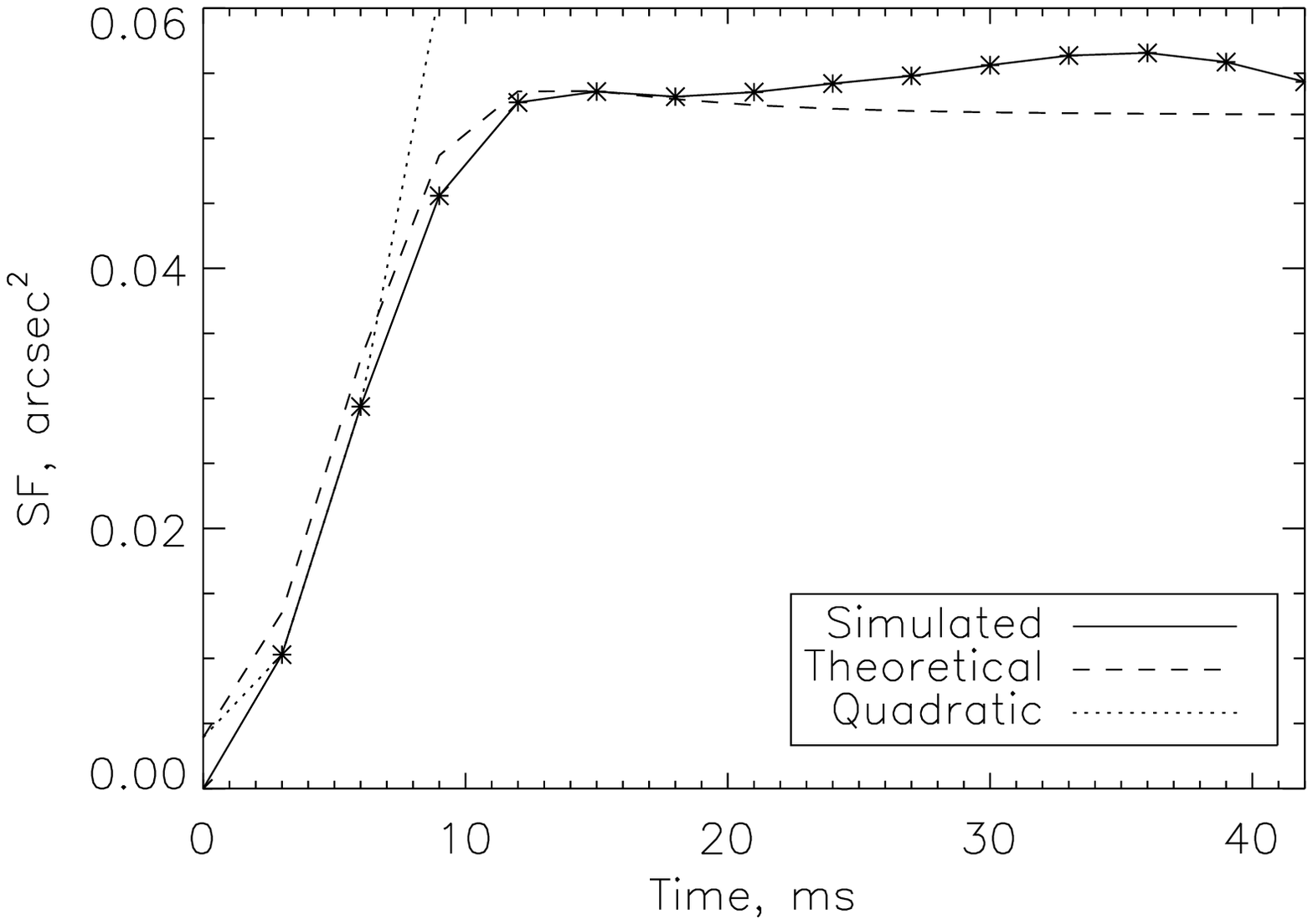}
 }
\caption{\label{fig:sf}  Temporal  structure  functions  of  simulated
  measurements of  the ring  radius for wind  speeds 10\,m/s  (left) and
  20\,m/s (right)  and $r_0  = 0.1$\,m seeing  (time constants  $t_1$ of
  3.36 and 1.68\,ms, respectively). }
\end{figure*}

To  measure  the interferometric  or  AO  time  constant, we  need  an
observable related to  $\overline{V}_2$ or $\overline{V}_{5/3}$.  The 
atmosphere  consists of  many layers  with different  wind  speeds and
directions, so  a true $C_n^2$-weighted  estimator (\ref{eq:Vp}) is
required. Its response should be independent of the wind direction.

Wavefront distortions are commonly decomposed into Zernike modes (Noll
\cite{Noll76}). The first mode, piston, cannot be sensed with a single
telescope and the  two  subsequent  modes,  tip and  tilt,  tend  to  be
corrupted by  telescope vibrations.  Of the remaining  modes, the next
three -- defocus and two astigmatisms -- have the highest variance and
are the best candidates for measuring atmospheric parameters.

The total turbulence integral (or  $r_0$) is typically measured by the
DIMM (Sarazin, \& Roddier \cite{DIMM}). 
Lopez (\cite{Lopez92}) tried  to derive  $\tau_0$ from the  speed of  the DIMM
signal, but this method did not  prove to be practical.  Because of its
intrinsic   asymmetry,  DIMM   does  not   provide  an   estimator  of
$\overline{V}$  that is  independent of  the wind  direction.   On the
other hand, the fourth Zernike mode (defocus) is rotationally symmetric.

We show in Appendix\,B that  the variance of defocus velocity provides an
estimator of  the time  constant $t_1$.  The  variance of  the defocus
itself gives a  measure of $r_0$. Thus, we can  measure both $r_0$ and
$\overline{V}_2$.   The method  is  based on  series  of fast-defocus
measurements, and we call it FADE (FAst DEfocus).
The details of the future FADE  instrument still need to be worked out
and  will be  a subject  of the  forthcoming paper.   
Here  we present
numerical simulations  to show the  feasibility of this  approach.  We
simulated a  telescope of $d =  0.35$\,m diameter with  a small central
obstruction $\epsilon = 0.1$. A conic aberration was introduced to form
ring-like images  (Fig.\,\ref{fig:movie}). This configuration resembles
a DIMM with  a continuous annular aperture. The  ring radius $3''$ was
chosen.

Monochromatic ($\lambda  =  500$\,nm) images  were  computed on  a
64$^2$ pixel  grid from the  interpolated distortions and  binned into
CCD pixels  of $0.86''$ size. We simulated  photon noise corresponding
to a star of $R=2$  magnitude and 3\,ms exposure time (20\,000 photons per
frame) and added a readout noise of 15 electrons rms in each pixel.

The radius $\rho$  of the ring image is calculated in  the same way as
standard  centroids,  by simply replacing  coordinate  with radius.   The
radius  fluctuations  $\Delta \rho$  serve  as  an  estimator for  the
defocus coefficient  $a_4$.  The radius change is  approximated by the
average slope of  the Zernike defocus between inner  and outer borders
of the aperture:
\begin{equation} 
\Delta  \rho = C_\rho \; a_4 \approx [ 2 \sqrt{3}  (1 +  \epsilon)/\pi \;
(\lambda/d) ]\; a_4 .
\label{eq:rho}
\end{equation}

The complex amplitude of the light distorted by two phase screens at 0
and 10\,km altitude with combined $r_0 = 0.1$\,m was pre-calculated on a
large square  grid (15\,m size, 0.015\,m pixels).   This distribution is
periodic  in  both coordinates,  and it  was  ``moved''  in front  of  the
aperture  in a  helical pattern  with the  wind speed  $V$  to simulate the
temporal evolution of  the wave-front.  The exposure time
 $\Delta t = 3$\,ms corresponds to a wave-front shift $V \Delta t = 0.06$\,m for $V
= 20$\,m/s, such that the initial quadratic part of the defocus SF ($\beta = 2
V t /d < 1$) extends only to $\sim 3 \Delta t$.

Figure\,\ref{fig:sf}  shows the  structure function, $D_\rho$, of  the ring-image
radius  calculated from several seconds of simulated data.  It
contains a small  additive component due to the  measurement noise (in
this case $0.05''$  rms), which was determined from  the data itself by
a quadratic fit  to the 2nd and 3rd points  and its extrapolation to
zero.  The dashed lines are the theoretical SFs of defocus computed by
(\ref{eq:D-4})  and  converted  into  radius  with  the  coefficient
$C_\rho$  (\ref{eq:rho}).   The slope  between  the  second and  third
points of the simulated SF closely matches the analytical formula.

To measure the speed of defocus  variations, it is sufficient to fit a
quadratic  approximation  to the  initial  part  of  the measured  SF,
$D_\rho(t) \approx a t^2$. Considering the noise, the best estimate of
the coefficient $a$ is obtained from  the second and third points, $ a
= [ D_\rho  (2 \Delta t) - D_\rho ( \Delta  t)]/(3 \Delta t^2)$.  This
estimator is not biased by white measurement noise.  Equating the
quadratic  fit  to the  theoretical  expression  $D_\rho(t) =  0.0269\,
(C_\rho\,t/t_1)^2$, we get a recipe for calculating the time constant
from the experimental data,
\begin{equation}
t_1 \approx 0.284\, C_\rho \Delta t \; 
[ D_\rho (2 \Delta t) -  D_\rho ( \Delta t)]^{-1/2} .  
\label{eq:t1meas}
\end{equation}
Application of this  formula to the simulated data  gives $t_1$ values
of 3.88 and  2.20\,ms for wind speeds 10 and  20\,m/s, while the input
values  are  3.36  and  1.68\,ms. Our  simulated  instrument  slightly
over-estimates $t_1$ because the chosen  exposure time of 3\,ms is too
long.   Indeed, the  error  gets worse  for  a higher  wind speed  and
disappears  for $V =  5$\,m/s (true  and measured  $t_1$ are  6.73 and
6.62\,ms) or  for a shorter exposure  time.  In the  real situation of a 
multi-layer atmosphere, the  experimental SF will be the sum of the SFs
produced by different layers. The  contribution to the ``jump'' of the
SF $ D_\rho (2  \Delta t) - D_\rho ( \Delta t)$  from fast layers will
be reduced (in comparison with the quadratic formula) and will cause a
bias in the measured $t_1$, increasing its value.

The crudeness of our simulations (discrete shifts of the phase screen,
approximate    $C_\rho$, etc.)      also    contributes    to    the
mismatch. Averaging of  the image during finite exposure  time has not
been simulated yet.   The response and bias of  a real instrument will
be studied  thoroughly by a  more detailed simulation.   However, the
feasibility of  the proposed technique for measuring  $t_1$ is already
clear.

The  next two  Zernike  modes number  5  and 6  (astigmatism) are  not
rotationally  symmetric.  
However,  the sum  of the  variances  of the
velocities  of two  astigmatism coefficients  is again  symmetric.  In
fact, it has the same spatial and temporal spectra as defocus, with a  twice  larger
variance.  Therefore,  simultaneous  measurement  of the two
astigmatism coefficients can be  used to estimate the atmospheric time
constant  in the  same way  as  defocus.  Other  measurables that  are
symmetric and have  a cutoff at high frequencies can  be used as well.
However,  defocus  and astigmatism  have the  largest  and slowest
atmospheric variances making it easier to  measure than other
higher-order modes.

The  FADE technique  can  be  applied in a straightforward  way to  the
analysis of  the AO  loop data,  as a simple  alternative to  the more
complicated method developed by Fusco et al. (\cite{Fusco2004}).

\section{Conclusions}

We  reviewed  the theory  of  fast  temporal  variations in  the  phase
difference  in  a  large-baseline  interferometer. For  a  practically
interesting case of  large apertures $d > r_0$,  the piston SF usually
exceeds  1  rad$^2$   at  the  aperture  crossing  time   $t  =  d/
\overline{V}$. Hence,  shorter times are of interest  where the piston
SF  is  quadratic  (rather  than  $ \propto  t^{5/3}$).  The  relevant
atmospheric  time constant  is $t_1$.  However, the  standard  AO time
constant  $\tau_0$  also provides  a  good  estimation  of the  piston
coherence  time.   Both these  parameters  essentially  depend on  the
turbulence-weighted average wind speed $\overline{V}$.

A brief review  of available methods for measuring  $\tau_0$ shows the
need for a  simple technique suitable for site  testing or monitoring,
i.e. working  on a small-aperture  telescope. The FAst  DEfocus (FADE)
method proposed  here fulfills  this need.  We  argue that, for  a given
aperture  size, this is  the best  way of  extracting the  information on
$\tau_0$.   The feasibility of  the method  is proven  by simulation,
which opens a way  to the development  of a real instrument.   An instrument
concept using a small telescope,  some simple optics, and a fast camera
will be described in a subsequent article.


\appendix

\section{Derivation of the piston structure function}

The spatial power  spectrum of the piston is  derived from the spatial
atmospheric phase spectrum (Roddier \cite{Roddier81})

\begin{eqnarray}
W_{\varphi}({\bf f}) = 
0.00969 \; k^2
\int_0^{+\infty} ( f^2 + L_0^{-2})^{-11/6} \;  C_n^2 \; {\rm d\/}h ,
\label{eq:Wphi}
\end{eqnarray}
where ${\bf  f}$ is the  spatial frequency, $L_0$ the turbulence outer
scale   at  height   $h$,  and the other  notations  were   introduced  in
Sect.\,\ref{sec:heuristic}.  We  drop the explicit  dependence of $C_n$,
$L_0$, and all following altitude dependent-parameters on $h$,
to ease  the  reading of  the formulae.   The
spatial filter  that converts  $W_{\varphi}({\bf f})$ into  the piston
power spectrum $W_\phi ({\bf f})$ is

\begin{eqnarray}
\label{eq:Mphi}
M^2({\bf f}) &=& [2 \; \sin ( \pi {\bf fB}) \; A({\bf f}) ]^2 \\
W_\phi ({\bf f}) &=& M^2({\bf f}) \; W_{\varphi}({\bf f}) ,
\end{eqnarray}
for  a baseline  vector ${\bf  B}$  and the  aperture filter  function
$A({\bf f})$.  For a circular aperture of diameter $d$, $A ({\bf f}) =
2 J_{1}(\pi fd)/  (\pi fd)$ and $f = | {\bf  f}|$. There $J_n$ stands
for the Bessel function of order $n$.

As usual,  we assume that  turbulent layers are transported  with wind
speed ${\bf  V\/}$ directed at an  angle $\alpha$ with  respect to the
baseline. The temporal  power spectrum of the piston  is then obtained
by integrating in the frequency plane  over a line displaced by $f_x =
\nu/V$ from the coordinate origin  and inclined at angle $\alpha$. Let
$f_y$  be  the  integration  variable  along  this  line  and  $f^2  =
f_x^2+f_y^2$.  The temporal spectrum  equals
\begin{eqnarray}
\label{eq:w-nu}
w_\phi (\nu) &=& \frac{1}{V} \int_{- \infty}^{+\infty} 
W_\phi \left( f_x \cos \alpha +  f_y \sin \alpha , f_y \cos \alpha  - f_x \sin \alpha
\right) \; {\rm d\/}f_y  \nonumber \\
&=& 0.0388 \; k^2 
\int_0^{+\infty} V^{-1} C_n^2  \;  {\rm d\/}h 
\int_{- \infty}^{+\infty} (f^2+L_0^{-2})^{- 11/6} \nonumber \\
&& \times \left[\sin \left( \pi B f_x \cos
\alpha + \pi B f_y \sin \alpha \right)  A (f)
\right] ^2 \; {\rm d\/}f_y .
\end{eqnarray}
We use the  rotational symmetry of the aperture  filter.  This formula
can be found in Conan  et al. (\cite{Conan95}) in a slightly different
form.  The temporal power spectrum  is defined here on $\nu = (-\infty,
+\infty)$ to keep the analogy with spatial power spectra.

The  temporal structure function of the piston is
\begin{eqnarray}
D_{\phi}   (t)   &=&\int_{-\infty}^{+\infty}   2[1-{\rm  cos}(2\pi   t   \nu)]
\; w_\phi(\nu) \; {\rm d}\nu .
\label{eq:Dphi4}
\end{eqnarray}
For an interferometer with a large baseline $B \gg d$, the width of the
aperture filter is much larger than the  period of the
$\sin^2$ factor in (\ref{eq:w-nu}). We can then replace the $\sin^2$
with its average value 0.5. Assuming also that $L_0 \gg d$, we obtain
an approximation for the piston power spectrum
\begin{eqnarray}
\label{eq:w-nu1}
w_\phi (\nu)& \approx & 0.0194 \; k^2
\int_0^{+\infty} V^{-1} \; C_n^2 \;  {\rm d\/}h  \nonumber \\
&&\times\; \int_{- \infty}^{+\infty}  A^2 (f) f^{- 11/3} \; {\rm d\/}f_y .
\end{eqnarray}
With this approximation,  
\begin{eqnarray}
D_{\phi}(t)&=& 0.0388 \;  k^2 
\int_0^{+\infty} C_n^2 \;  {\rm d\/}h \int\int_{- \infty}^{+\infty} [1-{\rm cos}(2\pi t f_x V)] \nonumber \\
&&\times\; A^2 (f) f^{- 11/3} \; {\rm d\/}f_x {\rm d\/}f_y   \nonumber \\
&=& 0.244 \;  k^2 
\int_0^{+\infty} C_n^2 \;  {\rm d\/}h \int_0^{+\infty} [ 1-J_0(2\pi tVf) ] \nonumber \\
&&\times\; A^2 (f)  f^{- 8/3} \; {\rm d\/}f .
\end{eqnarray}
We  used  the  relation  (Gradshteyn  \&  Ryzhik  \cite{Gradshteyn65}):
$\int_0^{2 \pi} \cos (2  \pi z \cos \theta) {\rm d} \theta  = 2 \pi \;
J_0 (2 \pi z)$.  For a circular aperture of diameter $d$,
 \begin{eqnarray}
D_{\phi} (t)&=& 1.641 \;  k^2 d^{5/3}
\int_0^{+\infty} C_n^2 \;  {\rm d\/}h 
\; K_1(2 t V/d) , 
\label{eq:D-K}
\end{eqnarray}
where the new dimensionless variables are $\beta=2tV/d$ and $x = \pi f
d$ and the function $K_1(\beta)$
 \begin{eqnarray}
K_1(\beta)&=& \int_0^{+\infty} [2 J_1(x)/x ]^2 x^{- 8/3} \; [1-J_0(\beta
  x)]\; {\rm d\/}x  \nonumber \\
 &\approx& 1.1183 \frac{\beta^2}{( 4.7 + \beta^2)^{1/6}}.
\label{eq:K1}
\end{eqnarray}
The  approximation  of $K_1(\beta)$  is  accurate to  1\% for  all
values of  the argument and  reproduces the analytic solutions  of the
integral for very large and very small $\beta$. For example, for large
$\beta$  the aperture filter tends to one; hence 

\begin{eqnarray} 
K_1(\beta) &\approx& \int_0^{\infty} x^{-8/3} \; [ 1 - J_0(\beta x)] \; {\rm d\/}x \nonumber \\
&=& \pi / [ 2^{8/3}\,\Gamma^2(11/6) \sin (5 \pi/6) ] \; \beta^{5/3} 
= 1.1183 \; \beta^{5/3} 
\end{eqnarray}
(cf. Eq. 20 in Noll \cite{Noll76}). It follows that for $t> d/ \overline{V} $
\begin{eqnarray}
D_{\phi} (t) &\approx& 13.77 \;  (\overline{V}_{5/3} \, t /r_0)^{5/3}
= (t/t_0)^{5/3} .
\end{eqnarray}
For $t <  d/ \overline{V} $, $K_1(\beta) \approx 0.864 \; \beta^2$ and
\begin{eqnarray}
D_{\phi} (t) &\approx& 13.41 \;  (\overline{V}_2\,t /r_0)^2 \;
(r_0/d)^{1/3} 
= (t/t_1)^2 .
\end{eqnarray}
We recover (\ref{eq:Dphi3}). This proves that  the initial part
of the  piston SF  is indeed defined  by the overall  wavefront tilts.

For a single turbulent layer, the piston SF is directly proportional
to $K_1(\beta)$. Considering the small difference between two
alternative definitions of the average wind speed, 
$ \overline{V}_{5/3} \approx  \overline{V}_{2} \approx  \overline{V}$,
a good approximation for the SF at all time increments will be

\begin{equation}
D_{\phi} (t) \approx 3.88 \; (d/r_0)^{5/3} \;  K_1(2 t \overline{V}/d).
\end{equation}
With   the   approximation    (\ref{eq:K1}),   we finally obtain  
(\ref{eq:Dphi-model}).

\section{Fast focus variation}

The temporal  power spectrum of the Zernike  defocus coefficient $a_4$
is given in Conan et al. (\cite{Conan95}) as
\begin{eqnarray}
w_4(\nu)
&=& 0.00969 \;  k^2 \; 
\int_{-\infty}^{+\infty} V^{-1} C_n^2 \; {\rm d\/}h \nonumber \\
&&\times \; \int_{- \infty}^{+\infty}  A_4^2(f) f^{-11/3} 
{\rm d\/}f_y ,
\end{eqnarray}
where $A_4(f)  = 2  \sqrt{3} J_3(\pi  f d)/(\pi f  d)$ is  the spatial
filter corresponding  to the defocus  on a clear aperture  of diameter
$d$ (Noll \cite{Noll76}), $f_x = \nu/V$,  $f^2 = f_x^2 + f_y^2$, and we
assume $L_0  \gg d$.  This expression is  similar to (\ref{eq:w-nu1})
but  has a  two times  smaller  coefficient and  a different  aperture
filter.
The variance of defocus is a function of the Fried parameter:
\begin{eqnarray}
\sigma_4^2 &=&
\int_{-\infty}^{+\infty} w_4(\nu) {\rm d}\nu  \nonumber \\
&=& 0.00969  \;  k^2 \; 
\int_0^{+\infty} C_n^2 {\rm d\/}h
\int \int_{-\infty}^{+\infty} A_4^2(f) \; f^{-11/3}  
{\rm d}f_x\,{\rm d}f_y \nonumber \\
&=& 0.0232 \; (d/r_0)^{5/3}.
\label{eq:sigma4-1}
\end{eqnarray}
The variance of  the defocus velocity has the  following dependence on
atmospheric parameters:
\begin{eqnarray}
S_4^2 &=&
\int_{-\infty}^{+\infty} (2\pi\nu)^2 w_4(\nu) {\rm d}\nu  \nonumber \\
&=& 0.383  \;  k^2 \; 
\int_0^{+\infty} V^2 C_n^2 {\rm d\/}h \nonumber \\
&& \times\; \int \int_{-\infty}^{+\infty} f_x^2 \;  A_4^2(f) \; f^{-11/3} 
{\rm d}f_x\,{\rm d}f_y .
\label{eq:S4-1}
\end{eqnarray}
We set $x = \pi f d$ and find:
 \begin{eqnarray}
S_4^2 & \approx & 9.858  \;  k^2 \;  d^{-1/3}
\int_0^{+\infty} V^2 C_n^2 {\rm d\/}h 
\int_0^{+\infty}  J_3^2(x) x^{-8/3} {\rm d}x  \nonumber  \\
   & = & 
0.360  \; (\overline{V}_2 /r_0 )^2 \; (r_0/d)^{1/3}  = 0.0269 \; t_1^{-2} .
\label{eq:S4}
\end{eqnarray}
The  transformation from (\ref{eq:S4-1})  to (\ref{eq:S4})  involves a
coefficient increase  by $ 12 \pi^{2/3}$, while  the definite integral
is  equal to  $ \Gamma(8/3)  \Gamma(13/6) /  [  2^{8/3} \Gamma^2(11/6)
\Gamma(29/6) ] = 0.01547$.

The SF of  defocus $D_4(t)$ is derived in analogy  with the piston SF,
replacing  the   response  $A_1(f)$  for  piston   with  $A_4(f)$  for
defocus. The coefficient is 2  times smaller because only one aperture
is considered. In analogy with (\ref{eq:D-K}), 
\begin{eqnarray}
D_{4} (t)&=& 0.821 \;  k^2 d^{5/3}
\int_0^{+\infty} C_n^2 \;  {\rm d\/}h 
\; K_4(2 t V/d) ,
\label{eq:D-4}
\end{eqnarray}
 \begin{eqnarray}
K_4(\beta)&=& 12\; \int_0^{+\infty} [J_3(x)/x ]^2 x^{- 8/3} \; [1-J_0(\beta
  x)]\; {\rm d\/}x \nonumber \\
 &\approx & \frac{0.0464\,\beta^2 + 0.024\,\beta^6 }{1 + 1.2\,\beta^2 + \beta^6}.
\label{eq:K4}
\end{eqnarray}
The  approximation has  a relative  error  less than  2\% and  correct
asymptotes.   Unlike $K_1$,  the $K_4$  function saturates  for large
arguments.  Considering only the  initial quadratic  part of  $K_4$ at
$\beta \ll 1$, we write for small time intervals
\begin{eqnarray}
D_{4} (t)& \approx & 0.360 \; (t \overline{V}_2 /r_0)^2 \; (r_0/d)^{1/3}
= 0.0269 \; (t/t_1)^2 .
\label{eq:D-4a}
\end{eqnarray}
%


%

\end{document}